\documentclass[twocolumn,showpacs,preprintnumbers,amsmath,amssymb]{revtex4}
\usepackage{graphicx}% Include figure files
\usepackage{amsmath}
\begin{document}

%\preprint{draft}

\title{Possible Phase Transition Deep Inside the Hidden Order Phase of Ultraclean URu$_2$Si$_2$}

\author{H.~Shishido$^\mathrm{1}$, K.~Hashimoto$^\mathrm{1}$, T.~Shibauchi$^\mathrm{1}$, T.~Sasaki$^\mathrm{2}$, H.~Oizumi$^\mathrm{2}$, N.~Kobayashi$^\mathrm{2}$, T.~Takamasu$^\mathrm{3}$, K.~Takehana$^\mathrm{3}$, Y.~Imanaka$^\mathrm{3}$, T.~D.~Matsuda$^\mathrm{4}$, Y.~Haga$^\mathrm{4}$, Y.~Onuki$^\mathrm{4,5}$,  and Y.~Matsuda$^\mathrm{1}$}

\affiliation{$^1$Department of Physics, Kyoto University, Kyoto 606-8502, Japan }%
\affiliation{$^2$2Institute for Materials Research, Tohoku University, Sendai 980-8577, Japan}%
\affiliation{$^3$National Institute for material Science,  Tsukuba, Ibaraki 305-0047, Japan}%
\affiliation{$^4$Advanced Science Research Center, Japan Atomic Energy Agency, Tokai 319-1195, Ibaraki, Japan}%
\affiliation{$^5$Graduate School of Science, Osaka University, Toyonaka, Osaka 560-0043, Japan}%

%\date{\today}

\begin{abstract}

To elucidate the underlying nature of the hidden order (HO) state in heavy-fermion compound URu$_2$Si$_2$, we measure electrical transport properties of ultraclean crystals in a high field/low temperature regime.  Unlike previous studies, the present system with much less impurity scattering resolves a distinct anomaly of the Hall resistivity at $H^{\ast}$=22.5~T well below the destruction field of the HO phase $\simeq$36~T.  In addition,  a novel quantum oscillation appears above a magnetic field slightly below $H^{\ast}$.  These results indicate an abrupt reconstruction of the Fermi surface, which implies a possible phase transition  well within the HO phase caused by a band-dependent destruction of the HO parameter. The present results definitely indicate that the HO transition should be described by an itinerant electron picture.

\end{abstract}

\pacs{71.27.+a,74.70.Tx,75.20.Hr,75.30.Mb}

\maketitle

The heavy fermion compound URu$_2$Si$_2$ has held attention of physicists for the past two decades owing to the presence of a `hidden order' (HO) transition at $T_c$=17.5~K \cite{Palstra,Maple,Schlabitz}.  Tiny magnetic moment appears ($M_0\sim 0.02 \mu_B$) below $T_c$ \cite{Bro91},  but $M_0$ is by far too small to explain the large entropy released during the transition.   Several exotic order parameters have been proposed for the HO phase\cite{OP}, which are based on either itinerant or localized pictures. However, the genuine HO parameter is still an open question.  Its elucidation is very important as it can lead to a discovery of new order parameters in strongly correlated electron systems.

Several remarkable features of the HO phase have been reported experimentally.  The electronic excitation gap is formed at a large portion of the Fermi surface (FS) below $T_c$ and most of the carriers ($\sim$90\%) disappears, resulting in a semimetallic state \cite{Sch87,Bel04,Beh05,Kas07}.  The remaining small number of carriers undergo a transition into an exotic superconducting state at $T_{sc}$=1.45~K \cite{Kas07,Oka08,Yan08}.  The neutron scattering experiments report the appearance of commensurate and incommensurate inelastic magnetic responses in the HO phase \cite{Bro91,Wie07,Vil08}.  The HO is affected by external parameters, such as pressure and magnetic field.   Above the critical pressure of $\sim$0.5~GPa, a true antiferromagnetic ordered state with large moments ($\sim 0.4\mu_B$/U) emerges \cite{Uem05,Ami07}, but no dramatic modification of the Fermi surface is reported between the HO and antiferromagnetic phases \cite{Has08,Nak03,Jef07,Ela09}.  In contrast, strong magnetic fields destroy the HO phase accompanying a radical reconstruction of the FS \cite{Jai02,Kim03,Lev08,Jo07,Oh07,Har03}.  Well below $T_c$, field-induced destruction of the HO occurs at $H_c\simeq$36~T.%+, followed by a cascade of phase transitions between the consecutive field-induced phases.  

Thus high magnetic field response is a clue for elucidating the HO parameter.     To date high field studies have been carried out by using crystals with rather low residual resistivity ratio ($RRR$).    Recently, extremely large magnetoresistance has been reported in the crystals with a very large $RRR(\sim 700)$ \cite{Kas07,Oka08}.  Such ultraclean crystals are expected to develop an advanced understanding about  the HO phase.  Here we present the  Hall effect, magnetoresistance and quantum oscillation measurements in the HO state of the ultraclean crystals.  We find several salient features which have never been reported in crystals with lower $RRR$ values, including a band-dependent destruction of the HO parameter and a possible transition deep inside the HO phase.

The URu$_2$Si$_2$ single crystals were grown by the Czochralski pulling method in a tetra-arc furnace \cite{crystal}.  The crystal used in this study displays a very low in-plane residual resistivity $\rho_0\simeq0.47~\mu\Omega$~cm and $RRR$ is 670 \cite{Kas07}.  The Hall effect and magnetoresistance were measured simultaneously up to 27~T in a transverse geometry for {\boldmath $J$}$\parallel a$ and for {\boldmath $H$} within the $bc$-plane.   %The single crystals were carefully examined by the X-ray diffraction and electron-probe microanalysis methods. We confirmed that the sample crystalizes in the ThCr$_2$Si$_2$-type structure.  The site occupancy and chemical composition were also consistent with ThCr2Si2 stoichiometry within an experimental accuracy of $<$1 \%.  %We obtained Hall resistivity from the transverse resistance by subtracting the positive and negative magnetic field data.
Figure 1(a)  displays the field dependence of the diagonal resistivity $\rho_{xx}$ well below $T_c$.  The magnetoresistance $\Delta\rho_{xx}(H)/\rho_{xx}(0)\equiv \frac{\rho_{xx}(H)-\rho_{xx}(0)}{\rho_{xx}(0)}$ is extremely large, $\sim$1200 at 25~T and $T$=0.11~K.  Here we obtained $\rho_{xx}(0)$  by the extrapolation of the normal state resisitivity in zero field above $T_{sc}$, assuming $\rho_{xx}(T)=\rho_0+AT^2$.  At very low temperatures, the Shubnikov-de~Haas~(SdH) oscillations are clearly seen.  As shown in the inset of Fig.~1(a),  $\rho_{xx}$ exhibits  nearly $H^2$-dependence in the low field regime.   No saturating behavior is observed for any field directions.  Such an exceptionally large magnetoresistance arises from the combination of the peculiar electronic structure, i.e. extremely high purity, very low carrier density, and nearly perfect compensation (equal number of electrons and holes) \cite{Kas07,Pippard}. 

%%%%%%%%%%%%%%%%%%%%%%%%%%%%%%%%%%%FIG 1%%%%%%%%%%%%%%%%%%%%
\begin{figure}[t]
\includegraphics[width=80mm]{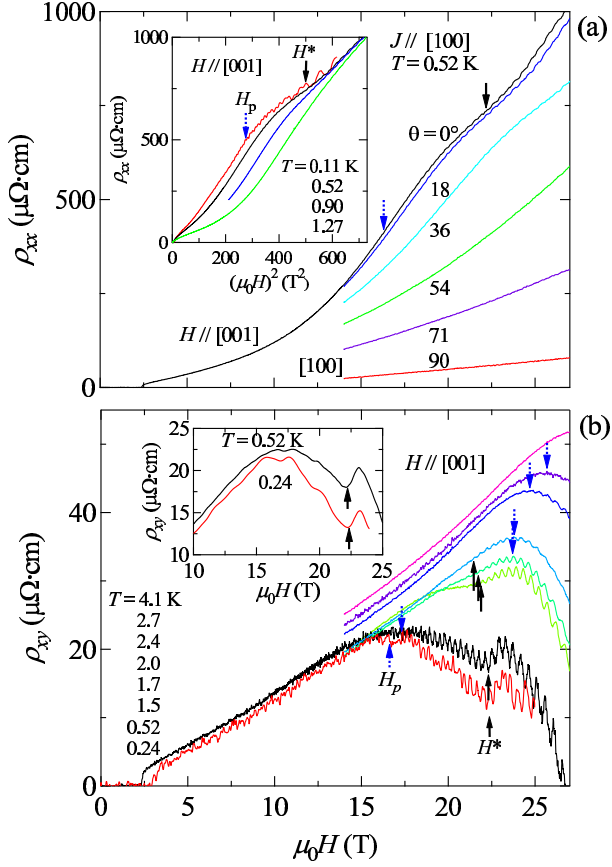}%
\caption{(color online) (a) $H$-dependence of $\rho_{xx}$ well below the HO transition  for several field orientations.  $\theta$ is the angle between {\boldmath $H$} and the $c$-axis.  Inset:  $\rho_{xx}$ plotted as a function of $H^2$.   (b) $H$-dependence of $\rho_{xy}$ at several temperatures for {\boldmath $H$}$\parallel c$.   The oscillation in $\rho_{xy}$ mainly arises from the $\varepsilon$ band.  Inset: Field dependence of $\rho_{xy}$ in which the oscillatory part is removed.  In each figure, the peak field $H_p$ and $H^{\ast}$ are marked by the blue dashed and black solid arrows, respectively.  %At $H^{\ast}$, $\rho_{xy}$ exhibits a steep increase and $\rho_{xx}$ has a infection point.   At $H_p$, $\rho_{xy}$ shows a peak and $\rho_{xx}$ deviates from $H^2$-dependence. 
}

\end{figure}
%%%%%%%%%%%%%%%%%%%%%%%%%%%%%%%%%%%FIG 1%%%%%%%%%%%%%%%%%%%%

The field dependence of the Hall resistivity  $\rho_{xy}$  is displayed in Fig.~1(b).  The quantum oscillations are also seen clearly.   There are two characteristic fields marked by $H_p$ (blue dashed arrows) and $H^{\ast}$ (black solid arrows).  With increasing $H$, $\rho_{xy}$ increases almost linearly and decreases after showing a maximum at $H_{p}$.  This broad peak of $\rho_{xy}$ is consistent with previous reports~\cite{Oh07}.     At $T$=0.52~K and 0.24~K, $\rho_{xy}$ exhibits another distinct anomaly at $H^{\ast}$=22.5~T.  This anomaly is less pronounced at elevated temperatures.  To see this more clearly, the inset of Fig.~1(b) displays $\rho_{xy}(H)$, in which the oscillatory part is removed.  At $H^{\ast}$, $\rho_{xy}$ exhibits a jump in a very narrow field range $(\Delta H \alt $ 1~T).   We stress that this distinct anomaly of $\rho_{xy}$ at $H^{\ast}$ is a characteristic feature of the ultraclean crystals, not reported previously.  These anomalies can also be seen in $\rho_{xx}$.  As shown in the inset of Fig.~1(a),  $\rho_{xx}$ deviates from $H^2$-dependence above $\simeq H_p$ and changes its slope again at $H^{\ast}$.

%%%%%%%%%%%%%%%%%%%%%%%%%%%%%%%%%%%FIG 2%%%%%%%%%%%%%%%%%%%%
\begin{figure}[t]
\includegraphics[width=80mm]{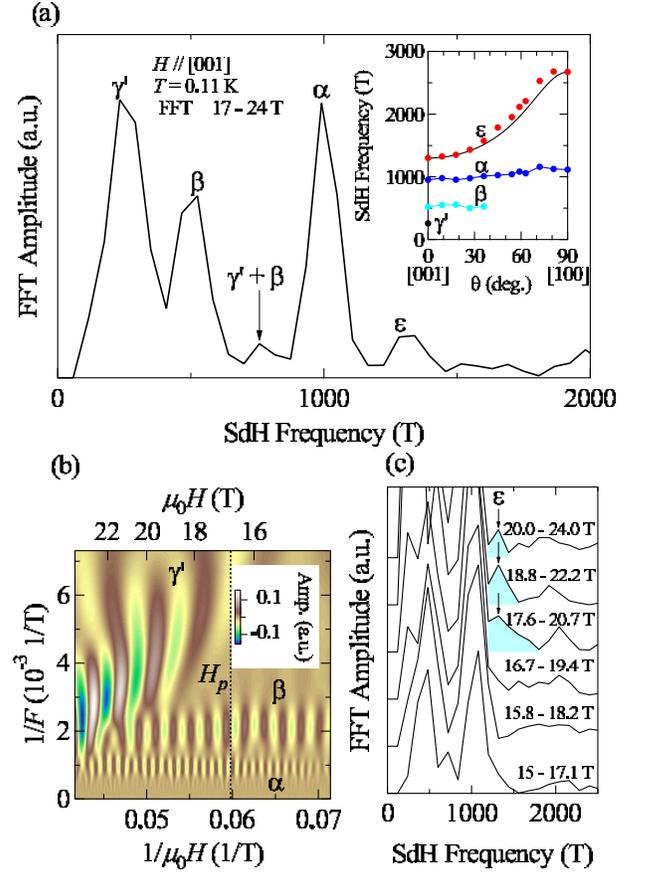}%
\caption{(color online)(a)FFT of the SdH oscillations in the field range 17-24~T for {\boldmath $H$}$\parallel c$ at $T$=0.11~K.  Inset: Angular variation of the SdH frequency for each branch. (b)  The power spectrum of the SdH oscillations (inverse of the frequency vs. $1/H$) displayed by the method of the continuous wavelet transform.  (c) FFT of the SdH oscillations in the several field interval.  Newly appeared $\varepsilon$-branch is shown by the hatched region.
}

\end{figure}
%%%%%%%%%%%%%%%%%%%%%%%%%%%%%%%%%%%FIG 2%%%%%%%%%%%%%%%%%%%%

Before analyzing the data, we briefly discuss the electronic structure in the low field regime of the HO phase.  Previous de Haas-van Alphen effect studies in fields up to 17~T resolved three frequencies for $\alpha$-, $\beta$- and $\gamma$- FS sheets, which correspond to $F_{\alpha}$=1.05~kT, $F_{\beta}$=0.42~kT and $F_{\gamma}$=0.19~kT with effective masses $m_{\alpha}$=13$m_0$, $m_{\beta}$=25$m_0$, and $m_{\gamma}$=8.2$m_0$, respectively, where $m_0$ is the free electron mass \cite{Nak03,Ohkuni,Ber97}.  Here, $F_i$ is proportional to the cross-sectional area $S_i$ of the $i$-th FS, $F_i=\hbar S_i/2\pi e$.   It has been shown that the largest $\alpha$-band is spherical and its volume corresponds to nearly 0.03 carriers per U-atom.  This value is close to the carrier number extracted from the Hall coefficient, indicating that the $\alpha$-band is a hole FS ($\rho_{xy}>0$) and governs the transport properties.  However the large Sommerfeld constant in the heat capacity($\sim 80$~mJ/K$^2$mol)~\cite{Maple} can not be accounted for the light $\alpha$-FS sheet, indicating the presence of missing heavy electron band \cite{Kas07}.

We now move on to the transport properties in the high field regime of the HO phase.   Figure 2(a) displays the fast Fourier transform (FFT) of the SdH oscillations observed in $\rho_{xx}$ for the field range between 17 and 24~T at $T=0.11$~K.   The branches which correspond to the $\alpha$- and $\beta$-FS sheets are observed.  The effective masses and the angular variation of the frequencies (inset of Fig.~2(a))  of these branches  are very close to those reported in the low field regime \cite{Nak03,Ohkuni,Ber97}.    A peak whose frequency is close to the $\gamma$-FS sheet in the low field regime, which is assigned as $\gamma'$, is also observed.  However the obtained effective mass $m_{\gamma'}= 27~m_0$ is much larger than the low field value.  Another branch of the SdH oscillations, $\delta$-branch with $F_{\delta}$=0.59~kT,  has been reported in Ref.\cite{Jo07} above 25~T.  However, this branch is not observed in the present crystals.  To examine the detailed field dependence of the frequency and amplitude of the SdH oscillations, we display in Fig.~2(b) the power spectrum of the oscillations yielded by means of the continuous wavelet transform.  It is notable that the $\gamma'$-branch emerges above $H_p$ and its frequency exhibits a large shift with $H$.

Another remarkable feature is the appearance of a new quantum oscillation, which is assigned as $\varepsilon$-FS sheet in Fig.~2(a).  As shown in the inset of Fig.~2(a), this $\varepsilon$-sheet is elliptical and its volume is nearly three times larger than the volume of the $\alpha$-sheet.  The effective mass of the $\varepsilon$-sheet is $m_{\varepsilon}=2.7m_0$, which is less than 30\% of $m_{\alpha}$.  Figure~2(c) displays the FFT spectrum obtained in the various field ranges.   While the $\varepsilon$-branch is not observable in the field range 16.7-19.4~T, it emerges in the range above 17.6-20.7~T (hatched area). Generally,  quantum oscillations arising from heavier bands decay faster with decreasing $H$.   Therefore, the fact that other branches with heavier masses are clearly observed in the whole $H$-range indicates that the low-field disappearance of the $\varepsilon$-branch with very light mass is not due to the decrease of the amplitude caused by the decrease of $H$.

Next we discuss the peculiar $H$-dependence of $\rho_{xy}$ characterized by $H_p$ and $H^{\ast}$.  A plausible explanation for the decrease of $\rho_{xy}$ above $H_p$, together with the concomitant deviation from the $H^2$-dependence of $\rho_{xx}$,  is a violation of the nearly perfect compensation realized in the low field regime.  This is supported by the continuous evolution of the FS across $H_p$, as evidenced by the development of $\gamma'$-FS sheet below $H_p$ shown in Fig.~2(b).   Such a FS evolution may come from the polarization of $\gamma'$-FS sheet, i.e. the shrinkage of one of the spin-split sheets to a point which occurs when the large Zeeman energy exceeds the Fermi energy \cite{Julian}.  In fact, the $\gamma'$-band with  a small volume and a large effective mass should have a small Fermi energy.  % We note that the peak of $\rho_{xy}$ at $H_p$ is not directly related to the appearance of the $\varepsilon$-branch.  This is because at $T$=0.11~K, the quantum oscillation due to the $\varepsilon$-FS sheet appears at 17.6-20.7~T as shown in Fig.~2(c), which is above $H_p \simeq 16~$T, while at elevated temperatures, the oscillation appears well below  $H_p$, as seen in Fig.~1(b).  

%%%%%%%%%%%%%%%%%%%%%%%%%%%%%%%%%%%FIG 3%%%%%%%%%%%%%%%%%%%%
\begin{figure}[t]
\includegraphics[width=80mm]{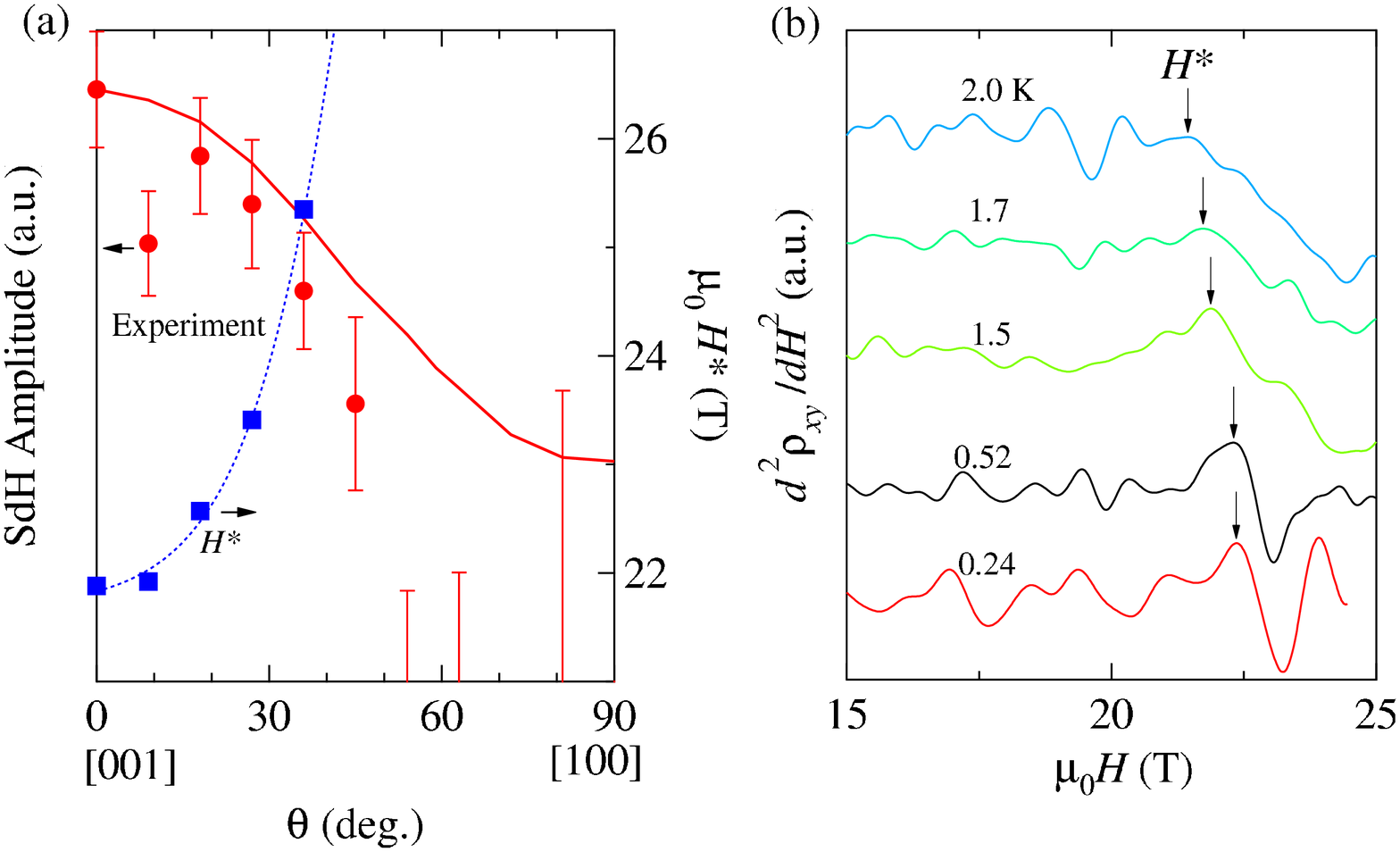}%
\caption{(color online) (a) Solid (blue) squares indicate the angular variation of $H^{\ast}$ determined by the inflection point of $\rho_{xx}(H)$, which is obtained by the field derivative of $\rho_{xx}$.  Dashed line is a guide for the eye.  Solid (red) circles indicate the angular dependence of SdH amplitude at $T$=0.52~K.  The solid line is the result of the Lifshitz and Kosevich formula. (b) Field dependence of the second derivative of $\rho_{xy}$ with respect to $H$.  The oscillatory part in $\rho_{xy}$ is filtered.  The arrows indicate $H^{\ast}$.
}
\end{figure}
%%%%%%%%%%%%%%%%%%%%%%%%%%%%%%%%%%%FIG 3%%%%%%%%%%%%%%%%%%%%

The transport anomaly at $H^{\ast}$ provides several pieces of important information for the electronic structure in the HO state.   We point out that the anomaly at $H^{\ast}$ is likely to be related to the appearance of the $\varepsilon$-FS sheet because of the following reason.  Figure~3(a) depicts the angular variations of $H^{\ast}$  (solid squares)  and the amplitude of SdH oscillations of the $\varepsilon$-FS sheet (solid circles) in the field range of 20 - 27~T.   $H^{\ast}$ increases steeply with increasing $\theta$.  The solid line is the SdH amplitude calculated from the Lifshitz and Kosevich formula \cite{LK}.   
%\begin{equation} 
%\frac{\tilde{\sigma}}{\bar{\sigma}} \propto H^{1/2} \left| \frac{ \partial^2 S_i}{\partial k^2} \right|^{-1/2} R_T  R_D \sin \left( \frac{2\pi F}{H} \right), 
%\end{equation}
%where $\tilde{\sigma}$ and $\bar{\sigma}$ are oscillatory and non-oscillatory parts of the conductivity, respectively. $ R_T = \frac{\lambda m_i T/H}{ \sinh ( \lambda  m_i T/H )} $ and $ R_D = \exp (-\lambda  m_i T_D/H )$ are the temperature and  Dingle  factors, respectively, where $\lambda = 2\pi^2 c k_\mathrm{B} /e\hbar$  and $T_D$ is the Dingle temperature \cite{spin}.  
The observed SdH amplitude decreases faster than the value expected in the Lifshitz and Kosevich formula at $\theta \agt 40^\circ$, where $H^{\ast}$ is out of the measurement range.  We note that  at $\theta \agt 40^\circ$ the $\varepsilon$-branch recovers in a higher field range of 24 - 27~T, indicating that the reduction in the field range of 20 -27~T is not due to the spin factor.  Therefore, the depression of the $\varepsilon$-branch in the field range of 20 -27~T indicates the correlation between $H^{\ast}$ and the SdH oscillation of the $\varepsilon$-branch.   The quantum oscillation of the $\varepsilon$-branch starts at fields slightly below $H^{\ast}$.  Such a precursor phenomenon is often observed in magnetic systems and in superconductors. As seen in Fig.~1(b), $\rho_{xy}(H)$ decreases more rapidly above $H^{\ast}$ than below $H^{\ast}$ and appears to become negative above 27~T at $T$=0.52~K.  Such a sign change indicates that the $\varepsilon$-FS is an electron pocket with high mobility.

It has been reported that new quantum oscillations often appear when the metamagnetic transition or magnetic breakdown occurs.  However, in either case, the jump of $\rho_{xy}$ has never been observed \cite{Julian,mbd}.    Therefore it is natural to attribute the observed twofold anomalies,  namely the jump of $\rho_{xy}$ and the appearence of the $\varepsilon$-FS sheet,  to an abrupt reconstruction of the FS, which is caused by a possible phase transition at $H^{\ast}$.  Since the electronic specific heat of the $\varepsilon$-band estimated from the SdH results is only $\gamma_{\varepsilon}\simeq 2.3$~mJ/K$^2$ mol, which is much less than the  Sommerfeld constant, very accurate thermodynamic measurements are required to detect this transition.

We emphasize that the Hall sign change as well as the jump of $\rho_{xy}$ is observed in the ultraclean crystals for the first time.  Therefore it is very likely that these anomalies originate from   the band-dependent nature of the HO parameter,  because each band manifests their characters when the interband scattering is strongly suppressed with high purity.  In such a case, each band has a different amplitude of the HO gap.  Then new bands can appear when the magnetic field exceeds the condensation energy of the smaller HO gap even well below  $H_c$,  at which the complete destruction of the HO state occurs.  This appearance of new bands would greatly change the properties of the electron orbits.   This situation bears striking resemblance to the multiband superconductors such as MgB$_2$, in which the amplitude of the superconducting order parameter is band dependent.  In this case, the superconductivity with the smaller gap is strongly suppressed at a ``virtual upper critical field" well below the true upper critical field \cite{Kas07,MgB}.  Note that the multiband superconducting nature is observable only in clean systems where the interband mixing is weak.

%%%%%%%%%%%%%%%%%%%%%%%%%%%%%%%%%%%FIG 4%%%%%%%%%%%%%%%%%%%%
\begin{figure}[t]
\begin{center}
\includegraphics[width=80mm]{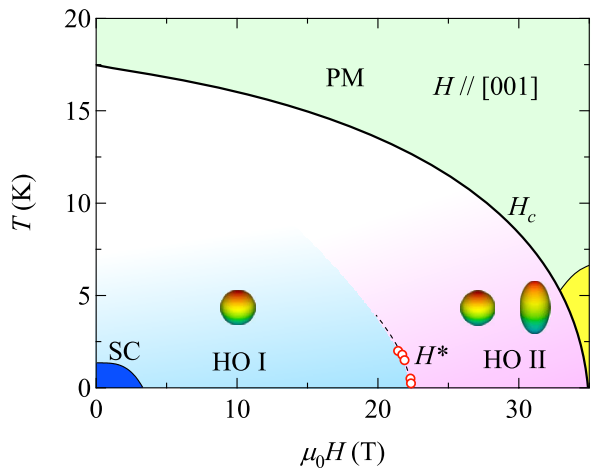}%
\caption{(color online) $H-T$ phase diagram for {\boldmath $H$}$\parallel c$.   SC and PM denote superconducting and paramagnetic regions respectively. The thick solid line is the critical field $H_c$ of the hidden order (HO) state \cite{Jai02,Kim03,Jo07,Oh07,Har03}.  The red circles are $H^{\ast}$ (see Fig.~3(b)), separating two distinguished HO state, HO~I and HO~II.  Spherical band sketches the (hole) $\alpha$-FS sheet.  HO~II is characterized by the appearence of the elliptical (electron) $\varepsilon$-FS sheet.    The high field region above $H_c$ shown by yellow is the field induced phase \cite{Jai02,Jo07,Oh07}.
}
\label{R_s}
\end{center}
\end{figure}
%%%%%%%%%%%%%%%%%%%%%%%%%%%%%%%%%%%FIG 4%%%%%%%%%%%%%%%%%%%%

Finally we discuss the $H-T$ phase diagram.  The temperature dependence of $H^{\ast}$ is determined by the kink positions in $d^2\rho_{xx}/dH^2$ vs. $H$ curves, shown in Fig.~3(b).   As the temperature is increased, $H^{\ast}$  decreases gradually.  Figure~4 displays the phase diagram obtained by the present study. % Unfortunately, it is hard to extrapolate the transition line to higher temperatures because thermal smearing hinders the anomalies at $H^{\ast}$ above $\sim 2$~K.  However, 
The results strongly suggest that the  HO phase contains two distinguished phases, HO~I and HO~II, each having different FS in the reciprocal space (see sketches in Fig.~4).   %It is intriguing that the energy scale of $H^{\ast}$ is close to the magnetic excitation gap at {\boldmath $Q$}$_C$ \cite{Bro91,Wie07,Vil08}.   

In summary, the transport studies by using ultraclean crystals of URu$_2$Si$_2$  reveal several outstanding features in the hidden order phase. Particularly, the observed transport anomalies and the appearance of a new quantum oscillation indicate an abrupt reconstruction of the Fermi surface.  This implies a possible phase transition well within the HO phase caused by a band-dependent destruction of the hidden order parameter. Our finding places a constraint on theories of the hidden order state, which has been discussed by localized or itinerant electron picture.  The band-dependent nature definitely indicates that the hidden order transition should be described by an itinerant picture.

We thank D.~Aoki, K.~Behnia, F.~Bourdarot, J.P.~Brison, J.~Flouquet, H.~Harima, E.~Hassinger, H.~Ikeda, G.~Knebel and  K.~Miyake for discussions.  %This work was supported by KAKENHI from JSPS and by Grant-in-Aid for the Global COE program ``The Next Generation of Physics, Spun from Universality and Emergence".  
%H.S. was supported by the JSPS Research Fellowship for Young Scientists.

\newpage


\begin{thebibliography}{99}
\bibitem{crystal}  The site occupancy and chemical composition of the crystals were  consistent with ThCr$_2$Si$_2$ stoichiometry within an experimental accuracy of $<$1 \%
\end{thebibliography}
\end{document}